\documentstyle[epsf,amssymb,amsmath]{mn}
\newif\ifAMStwofonts
\AMStwofontstrue

\ifoldfss
  \ifCUPmtlplainloaded \else
    \NewTextAlphabet{textbfit} {cmbxti10} {}
    \NewTextAlphabet{textbfss} {cmssbx10} {}
    \NewMathAlphabet{mathbfit} {cmbxti10} {} % for math mode
    \NewMathAlphabet{mathbfss} {cmssbx10} {} %  "   "    "
  \fi
  \ifAMStwofonts
    \ifCUPmtlplainloaded \else
      \NewSymbolFont{upmath} {eurm10}
      \NewSymbolFont{AMSa} {msam10}
      \NewMathSymbol{\upi}     {0}{upmath}{19}
      \NewMathSymbol{\umu}     {0}{upmath}{16}
      \NewMathSymbol{\upartial}{0}{upmath}{40}
      \NewMathSymbol{\leqslant}{3}{AMSa}{36}
      \NewMathSymbol{\geqslant}{3}{AMSa}{3E}

       \let\le=\leqslant
       \let\ge=\geqslant
    \fi
  \fi
\fi % End of OFSS

\ifnfssone
  \newmathalphabet{\mathit}
  \addtoversion{normal}{\mathit}{cmr}{m}{it}
  \addtoversion{bold}{\mathit}{cmr}{bx}{it}
  \newmathalphabet{\mathbfit} % math mode version of \textbfit{..}
  \addtoversion{normal}{\mathbfit}{cmr}{bx}{it}
  \addtoversion{bold}{\mathbfit}{cmr}{bx}{it}
  \newmathalphabet{\mathbfss} % math mode version of \textbfss{..}
  \addtoversion{normal}{\mathbfss}{cmss}{bx}{n}
  \addtoversion{bold}{\mathbfss}{cmss}{bx}{n}
  \ifAMStwofonts
    \ifCUPmtlplainloaded \else
      \UseAMStwoboldmath
      \makeatletter
      \new@mathgroup\upmath@group
      \define@mathgroup\mv@normal\upmath@group{eur}{m}{n}
      \define@mathgroup\mv@bold\upmath@group{eur}{b}{n}
      \edef\UPM{\hexnumber\upmath@group}
      \new@mathgroup\amsa@group
      \define@mathgroup\mv@normal\amsa@group{msa}{m}{n}
      \define@mathgroup\mv@bold\amsa@group{msa}{m}{n}
      \edef\AMSa{\hexnumber\amsa@group}
      \makeatother
      \mathchardef\upi="0\UPM19
      \mathchardef\umu="0\UPM16
      \mathchardef\upartial="0\UPM40
      \mathchardef\leqslant="3\AMSa36
      \mathchardef\geqslant="3\AMSa3E

       \let\le=\leqslant
       \let\ge=\geqslant
    \fi
  \fi
\fi % End of NFSS release 1

\ifnfsstwo
  \DeclareMathAlphabet{\mathbfit}{OT1}{cmr}{bx}{it}
  \SetMathAlphabet\mathbfit{bold}{OT1}{cmr}{bx}{it}
  \DeclareMathAlphabet{\mathbfss}{OT1}{cmss}{bx}{n}
  \SetMathAlphabet\mathbfss{bold}{OT1}{cmss}{bx}{n}
  \ifAMStwofonts
    \ifCUPmtlplainloaded \else
      \DeclareSymbolFont{UPM}{U}{eur}{m}{n}
      \SetSymbolFont{UPM}{bold}{U}{eur}{b}{n}
      \DeclareSymbolFont{AMSa}{U}{msa}{m}{n}
      \DeclareMathSymbol{\upi}{0}{UPM}{"19}
      \DeclareMathSymbol{\umu}{0}{UPM}{"16}
      \DeclareMathSymbol{\upartial}{0}{UPM}{"40}
      \DeclareMathSymbol{\leqslant}{3}{AMSa}{"36}
      \DeclareMathSymbol{\geqslant}{3}{AMSa}{"3E}

       \let\le=\leqslant
       \let\ge=\geqslant
    \fi
  \fi
\fi % End of NFSS release 2

\ifCUPmtlplainloaded \else
  \ifAMStwofonts \else % If no AMS fonts
    \def\upi{\pi}
    \def\umu{\mu}
    \def\upartial{\partial}
  \fi
\fi

\newcommand{\<}{\left<}
\renewcommand{\>}{\right>}	
\newcommand{\NG}{\rm NG}
\newcommand{\NL}{\rm NL}

\renewcommand{\min}{\rm min}
\renewcommand{\max}{\rm max}
\title{Probability of the most massive cluster under non-Gaussian initial conditions}
\author[] {
Laura Cay\'on$^{1}$, Christopher Gordon$^{2}$, Joseph Silk$^{2}$\\
1. Department of Physics. Purdue University. 525 Northwestern Avenue, West Lafayette, IN 47907-2036, USA\\
2. Oxford Astrophysics, Physics, DWB, Keble Road, Oxford, OX1 3RH, UK}
\date{\today}

\pagerange{\pageref{firstpage}--\pageref{lastpage}}
\pubyear{2010}

\begin{document}

\maketitle

\label{firstpage}

\begin{abstract}

\noindent  Very massive high redshift clusters can be used to constrain and test the  $\Lambda$CDM model. Taking into account the observational constraints of Jee et al. (2009) we have calculated the probability for the {\em most massive cluster\/} to be found in the range $(5.2-7.6)\times 10^{14}M_{\sun}$, between redshifts $1.4\le z \le 2.2$,
with a  sky area of 11 deg$^2$ and under non-Gaussian initial conditions. Clusters constrain the non-Gaussianity on smaller scales than current cosmic microwave background or halo bias data and so can be used to test for running of the non-Gaussianity parameter $f_{\NL}$. 
%Combining with WMAP7 data, we find that on cluster scales there is a 94\% probability for $f_{\NL}>0$. 

\end{abstract}


\section{Introduction}

In this article we use galaxy cluster observations to place constraints on primordial non-Gaussinaity.
The weak-lensing analysis of the galaxy cluster  XMMU J$2235.3-2557$ (hereafter  XMM2235) presented by Jee et al. (2009) indicates the existence of a massive cluster which 
has a mass  of
$M_{324}=(6.4 \pm 1.2)\times 10^{14}M_{\sun}$ at a redshift $z\approx1.4$.\footnote{The subscript indicates that the halo is defined as a sphere with a density which is 324 times the mean matter density of the Universe. Also, all errors will be given at the one standard deviation confidence intervals unless stated otherwise.}  The probability of observing a cluster with mass equal to or above the observed one is of the order of $5\times 10^{-3}$ for the surveyed area (11 deg$^2$) and redshift range $1.4\le z\le 2.2$. 
%Such a cluster would therefore appear to be
%almost a $3\sigma$ anomaly under the assumptions of the standard $\Lambda$CDM %model. 

The expected number of high redshift clusters is derived from the distribution of the primordial matter density fluctuations. For canonical single field inflation models the primordial matter density fluctuations are effectively Gaussian distributed. However, various alternatives have been proposed in which different amounts of non-Gaussianity are allowed (see for example Komatsu et al. (2009) for a review). Deviations from Gaussianity are often parametrized by the dimensionless non-linear coupling parameter $f_{\NL}$ (Komatsu \& Spergel 2001) through the `local' relationship
\begin{equation}
\Phi(\bf {x})=\phi(\bf{x})+f_{\NL}(\phi(\bf{x})^2-\<\phi(\bf{x})^2\>),
\label{fnl}
\end{equation}   
where $\Phi$ denotes the Bardeen gravitational potential and $\phi$ represents an isotropic Gaussian random field. Several works have calculated the effect of $f_{\NL}\ne 0$ on the predicted cluster number counts (Matarrese et al. 2000, LoVerde et al. 2008, Grossi et al. 2009, Desjacques, Seljak \& Iliev 2009, Desjacques \& Seljak 2010). It has been pointed out by Jimenez \& Verde (2009) and Sartoris et al. (2010) that for cluster XMM2235 not to be a rare event, values of $f_{\NL}$ above the current limits set by CMB observations might be required. Sartoris et al. (2010) included a discussion on the degeneracy between $f_{\NL}$ and $\sigma_8$. 
Conclusions from both works are based on the expected number of clusters with masses   {\em equal to and above\/} $5\times 10^{14} M_{\sun}$ at the observed redshift.

Holz \& Perlmutter (2010) proposed evaluating the probability of the most massive observed cluster in a survey. 
As the most massive cluster will be one of the easiest to find in the survey, it will be
 less sensitive to selection effects.
  They find that  XMM2235 is near the 3 sigma error contours in maximum mass and redshift for  a 11 deg$^2$ survey. Note that 
  assuming XMM2235 is the most massive cluster in the survey is conservative in that if it turns out there is a more massive one in that survey the probability would be even lower. In this article, we extend Holz \& Perlmutter's approach by evaluating how the {\em probability of the most massive cluster\/} is affected by non-Gaussianity.
  
The rest of this article is organized as follows. The theoretical background for calculating the expected number of counts in the presence of non-Gaussianity is presented in Section 1.  A discussion on how the probability distributions are built and what are the results obtained for the observed range of masses and redshifts is given in Section 2. The conclusions are given in  Section 3.

%\setcounter{figure}{0}
%\begin{figure*}
% \epsfxsize=95mm
% \epsffile{plot_fnl_number_clusters_WMAP7_2nongenh.eps}
% \caption{Expected number of clusters in a 11 deg$^2$ survey with mass in the range $5.2\times 10^{14}M_{\sun}\le M_{324}\le7.6\times 10^{14}M_{\sun}$ and a redshift in the range  $1.4\le z \le 2.2$ as a function of the level of non-Gaussianity parametrized by $f_{\NL}$. The solid line corresponds to the non-Gaussianity enhancement as calculated by LoVerde et al. (2008). The dashed line denotes the values considering the non-Gaussianity enhancement as derived by Matarrese et al. (2000). Results are calculated based on WMAP7 cosmology (Larson et al. 2010).}
% \label{f1}
%\end{figure*}

\section{Number counts in the presence of non-Gaussianity}

Based on the Press-Schechter formalism, Matarrese et al. (2000) and LoVerde et al. (2008) computed analytical expressions of the mass function $dn_{\NG}/dM$ in the presence of non-Gaussian initial conditions. These expressions have been tested against simulations (Grossi et al. 2009) for masses up to $\sim 10^{15} M_{\sun}$ and $f_{\NL}\sim 100$. They indicate that the Matarrese et al. (2000) approach produces a better fit at high masses/larger $f_{\NL}$ (also see G.~D'Amico et al. 2010, Maggiore \& Riotto 2009). For this reason we use the Matarrese et al. (2000) approach for modeling the effects of non-Gaussianity, although we use it somewhat beyond the ranges it has been tested for by N-body simulations.

One can write the non-Gaussian mass function in terms of the Gaussian one $dn/dM$ multiplied by a correction factor $R_{\NG}$

$$ {{dn_{\NG}(M,z,f_{\NL})}\over {dM}}={{dn(M,z)}\over {dM}}R_{\NG}(M,z,f_{\NL}))=$$
$${{\rho_{m}}\over {M}}{{d\ln\sigma(M,z)^{-1}}\over {dM}}f(\sigma(M,z))R_{\NG}(M,z,f_{\NL}),$$

where $\rho_{m}=\Omega_m \rho_c$, $\sigma^2(M,z)$ is the variance of the smoothed density fluctuations and $f(\sigma(M,z))$ is the mass function.  We assume non-Gaussianity deviations of the `local' form and adopt the CMB convention in the definition of $f_{\NL}$. The suggested barrier factor of $0.75$ is introduced in the computation of the critical collapse density to account for deviations from spherical collapse (Grossi et al. 2009). $R_{NG}$ increases with the skewness of the density distribution which increase for larger values of $f_{\NL}$. 
The above expression depends as well on the transfer function for which we use the
 fit by  Bardeen et al. (1995) with the correction introduced by Sugiyama (1995). We checked that 
the predicted number of clusters is not sufficiently different when the more accurate Eisenstein \& Hu (1998)
formula is used for our results to be significantly affected.

We are ultimately interested in calculating the probability density function of the most massive cluster within certain mass and redshift ranges. Such a calculation is based on the predicted average number of clusters. The average number of clusters per redshift interval $dN/dz$ in the mass range $[M_{\min},M_{\max}]$ is given by 
\begin{equation}
{{dN(z,f_{\NL})}\over {dz}}=f_{\rm sky}{{dV(z)}\over {dz}}\int_{M_{\min}}^{M_{\max}}dM {{dn_{\NG}(M,z,f_{\NL})}\over {dM}},
\label{dNdz}
\end{equation}
where $f_{\rm sky}$ represents the fraction of sky observed, $dV(z)/dz$ is the volume element given by
\[
\frac{dV}{dz}=\frac{4\pi}{H(z)}\left[\int_0^z \frac{dz'}{ H(z')}\right]^2
\]
and $H(z)$ is the Hubble parameter.

\section{Probability of the most massive cluster under non-Gaussian initial conditions}

 %Simulations of the number of clusters at a certain redshift interval for a given mass bin have been computed as Poisson realizations with an expected average number given by equation (\ref{dNdz}) integrated in redshift. For our purposes, only three mass bins where required: $M_{324}< 5.2\times10^{14}$, $ 5.2\times10^{14}\le M_{324}\le 7.6\times10^{{14}}$, and $M_{324}\ge 7.6\times 10^{14}$.
 %For a given $f_{\NL}$, the distribution of the  mass of the most massive cluster was obtained by determining the highest mass bin with  a non-zero number of clusters in each simulation.   

\begin{figure}
 \epsfxsize=9cm
 \epsffile{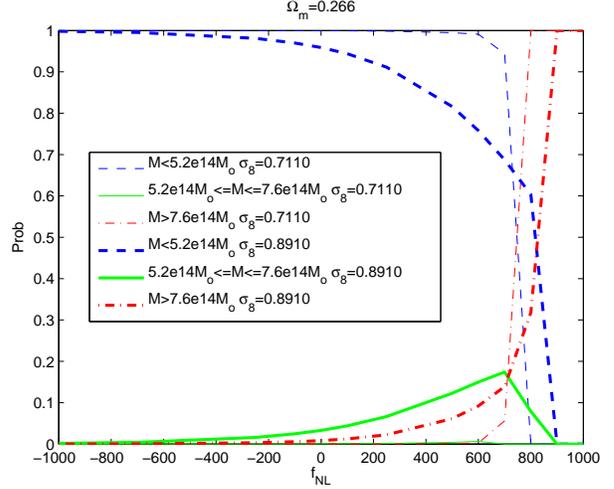}
 \caption{Variation of the probability of the most massive cluster, for different mass bins,
 as a function of non-Gaussianity ($f_{\NL}$) and $\sigma_{8}$. The blue (dashed), green (solid) and red (dash-dot) curves are different mass bins and the thick
and thin curves are different $\sigma_8$ values.
 \label{probability}}
\end{figure}

We were  interested in answering the following question: What is the probability of galaxy cluster XMM2235 being the most massive cluster within the surveyed volume? Jee et al. (2009) estimated the mass of this cluster to be $M_{324}=(6.4\pm 1.2)\times 10^{14} M_{\sun}$. We use the $\Lambda$CDM Jenkins et al. (2001) mass function
fitted for the 324 overdensity halo definition. 
We discuss the effect of mass function choice in Section~\ref{conclusions}.
 The redshift interval of the survey was $1.4\le z\le 2.2$ and the area 11 deg$^2$ ($f_{sky}$). We took as our fiducial cosmology, the best fit WMAP7 cosmological parameters (Larson et al. 2010): $h=0.710$; $\Omega_m=0.266$; $\Omega_b h^2=0.02258$; $n_s=0.963$; $\sigma_8=0.801$. We had three mass bins:
$M_{324} <5.2\times 10^{14}$, $5.2\times 10^{14}\le M_{324}\le 7.6\times 10^{14}$, and 
$M_{324}> 7.6\times 10^{14}$. 
By making our bin size of order or larger than the error bars of the observed mass estimate, we approximately account for the effects of Eddington bias (Mortonson et al.\ 2010).
We assumed the number of clusters in each mass bin was Poisson distributed with the expected number of clusters given by Eq.~\ref{dNdz} integrated over the redshift range of the survey. We then simulated a large number of surveys and evaluated how many times each mass bin contained the most massive
cluster of the survey.
  The probability of the most massive cluster being in each mass bin is then 
worked out by dividing by the total number of surveys simulated.
In Fig.~\ref{probability} we show how the probability of the largest cluster varies with
non-Gausianity and $\sigma_{8}$.
As would be expected, large non-Gaussianity and large $\sigma_{8}$ lead to larger most massive clusters being more probable (Desjacques et al. 2009, Desjacques \& Seljak 2010, Sartoris et al. 2010). This means that too much non-Gaussinity or too high a value of $\sigma_{8}$ is also disfavored as then the most massive cluster is more likely to be even larger than XMM2235. This is another aspect that distinguishes our approach from that adopted by Jee et al. (2009), Jimenez and Verde (2009) and Sartoris et al. (2010).  They looked at the probability of getting a more massive cluster than  XMM2235. In their approach, increasing the non-Gaussianity or $\sigma_{8}$ will always lead to  an improvement in the probability. 

We found that our results were very insensitive to the other cosmological parameters in the range of variation allowed by WMAP7 and so we kept them fixed at their WMAP7 maximum likelihood values, except for $\sigma_{8}$.
However, it should  be noted that the mass estimate of (Jee et al. 2009)
depended on the assumed concentration of the cluster. They used a concentration 
parameter (estimated from the N-body simulations of Gao et al. (2008)) of $c=3.20\pm 0.75$. The concentration parameter can be related to the halo formation redshift ($z_{f}$) which 
for a halo of mass ($M$), observed at redshift ($z_{o}$), can be  defined as 
the redshift that a fraction $F$ of the halo was in 
halos of mass greater than ($fM$) where $f$ and $F$ are positive constants less than one (Lacey and Cole 1993). Using the extended Press-Schechter formalism (Bond et al. 1991), this condition can be expressed as 
\begin{equation}
{\rm erfc}\left[  \frac{\omega(z_{f})-\omega(z_{o})}{\sqrt{2(\sigma(f M)^{2}-\sigma(M)^2)}}\right] =F  
\label{zf}
\end{equation}
where $\omega(z)$ is the growth function modified threshold overdensity of spherical collapse at redshift $z$ and 
can be approximated by (Lacey \& Cole 1993; Nakamura \& Suto 1997)
\begin{equation}
\omega(z) D(z) = 3\times \frac{(12\times \pi )^{2/3}}{20}\times \left(1-0.0123\times \log_{10}\left[1+x^3\right]\right)
\end{equation}
where $x=\frac{\left(\Omega _m{}^{-1}-1\right){}^{1/3}}{1+z}$ and $D(z)$ is the linear theory growth function normalized to equal one at redshift zero.
The characteristic halo density is related to the concentration parameter by
\begin{equation}
\delta_{c}=\frac{200}{3}\times \frac{c^3}{\log[1+c]-c/(1+c)}\, .
\end{equation}
Navarro et al. (1996) proposed that $\delta_{c}$ was
 proportional to
the mean density of the Universe at the time of the halo formation.
This model was re-calibrated by Gao et al. (2008) who found a good fit (for flat cosmologies) was given by setting $f=0.01$ and $F=0.1$ in Eq.~\ref{zf} and setting
\begin{equation}
\delta_{c}=600{\Omega_{m}(1+z_{f})^{3} \over (1-\Omega_{m})+ \Omega_{m}(1+z_{o})^{3}}\, .
\end{equation}
The primary dependence of $c$ on the cosmological parameters is due to the dependence of $\sigma$ in Eq.~\ref{zf} on $\sigma_{8}$. For our parameter choices we find that ${\partial c \over \partial \sigma_{8}}\approx 2$ and so the WMAP7+BAO+H0 prior we use for $\sigma_{8}$ (standard deviation of 0.024) leads to an uncertainty of in $c$ of about  0.048. 
Additionally, the Gao et al.\ (2008) simulations are based on a $\sigma_{8}=0.9$ cosmology, while the WMAP7+BAO+H0 prior we use has a mean of 0.809. This leads to an error of 0.2 in $c$.
However, this is negligible compared to the N-body fitting error for $c$ of 0.75 used by Jee et al. (2009) to estimate the mass error of XMM2235.
 So, within the tight constraints of WMAP7+BAO+H0 we can neglect the dependence of $c$ on the cosmological parameters. Hence our likelihood for XMM2235 alone will not be strictly correct outside the ranges of the WMAP7+BAO+H0 prior that we use. However, this will not affect our final constraints as they will be based on the posterior probability which includes the WMAP7+BAO+H0 prior.
 
Another systematic that should be considered is the effect of  contributions
from the triaxial shapes of cluster-sized halos and uncorrelated large-scale matter projections along the line-of-sight.
 This generally leads to an additional mass error of about 20\%
(Becker and Kravtsov (2010)). To account for this we added this error in quadrature to the error quoted by Jee et al. (2009). This widened our central bin width from 2.4 to $3.5 \times 10^{14} M_{\sun}$. We found this increased bin width had a negligible impact on our results.
%However, the current X-ray result does not appear to support any extreme
%line-of-sight elongation (Jee et al.\  2009, Jee 2011).
%If the weak-lensing mass is substantially boosted by the line-of-sight
%elongation, the X-ray mass should
%be considerably lower than the lensing mass. To the contrary, the
%Chandra mass estimate of the
%cluster is in fact slightly higher. In addition, by the same token,
%the extreme line-of-sight elongation would
%inflate the velocity dispersion significantly, which is not the case
%for XMM2235  (Jee et al.\  2009, Jee 2011).
 
\begin{figure}
 \epsfxsize=8cm
 \epsffile{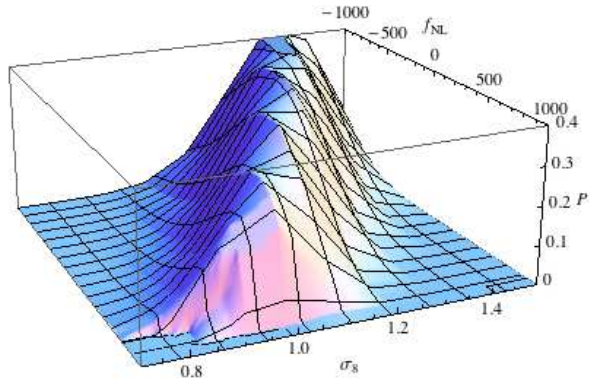}
 \caption{Likelihood ($P$) that XMM2235 is the most massive cluster
 as a function of non-Gaussianity ($f_{\NL}$) and $\sigma_{8}$.
 \label{likelihood}}
\end{figure}

 We generated a grid  of probabilities in $\sigma_{8}$ and $f_{\NL}$ for the XMM2235 being the maximum mass cluster. We then interpolated this grid to get our likelihood function, $p( \mbox{XMM2235 most massive} | f_{\NL}, \sigma_{8})$, which is plotted in Fig.~\ref{likelihood}.
As can be seen there is a strong degeneracy between $f_{\NL}$ and $\sigma_{8}$ as expected from Fig.~\ref{probability} and the above discussion.

\begin{figure}
 \epsfxsize=8cm
 \epsffile{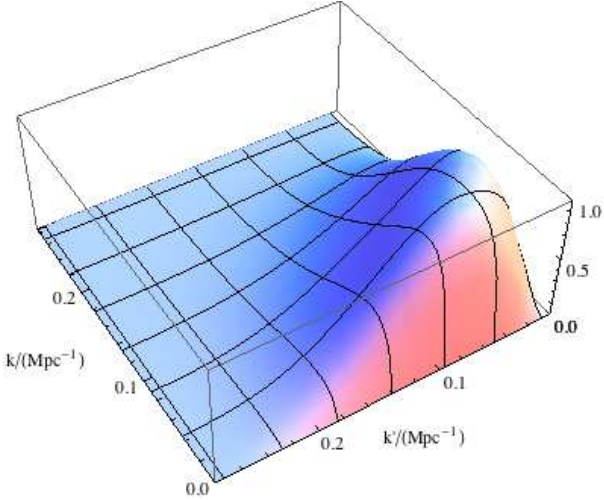}
 \caption{The kernel of the integral for the skewness of the linear density field smoothed on scales of 
 $16$ Mpc$^{-1}$.
 The angle between the wave-numbers $k$ and $k'$ has been integrated over.
  \label{kernel}}
\end{figure}

% we have calculated the probability of such a cluster, being the most massive in the survey, as a function of $f_{\NL}$. In discussing the probability of finding a cluster such XMM2235 being the most massive, one has to take into account the    
%well known degeneracy between $f_{\NL}$ and $\sigma_8$ (Desjacques et al. 2009, Desjacques \& Seljak 2010, Sartoris et al. 2010). 
 The WMAP7  data  constrains the local non-Gaussianity parameter to $-10<f_{\NL}<74$ at the $95\%$ CL (Komatsu et al. 2010). But this non-Gaussianity constraint is at wave-numbers of about $0.04$~Mpc$^{-1}$  (Komatsu \& Spergel 2010, LoVerde et al. 2008, Sefusatti et al. 2009).
 Non-Gaussianity has also been constrained to similar levels and  scales as WMAP7 by halo bias (Slosar et al. 2008). 
 The non-Gaussianity enhancement, $R_{\rm NG}$ depends on $f_{\rm \NL}$ only through its effect on the skewness of the smoothed density field (Matarrese et al.\ 2000). The skewness depends on an integral over two wave numbers, $k$ and $k'$, and the angle between them. We integrate over the angle and plot  the kernel in Fig.~\ref{kernel} for the density field smoothed on scales corresponding to the mass of XMM2235.
As can be seen the mean of the kernel is at around $k\approx 0.1$ Mpc$^{-1}$. Also 
the amplitude of the kernel remain non-negligable up to about 
$k \approx 0.2$ Mpc$^{-1}$. Therefore, the XMM2235 measurement probes smaller length scales than  the WMAP7 measurement of $f_{\rm \NL}$.
 There are multi-field models which produce a scale dependent $f_{\rm \NL}$ with the spectral index generally being first order in the slow roll parameters, but larger values are possible (see for example, Byrnes et al. 2009). Also,
 there are models which produce a step-function like change in $f_{\rm \NL}$ (Riotto and Sloth, 2010), so 
 $f_{\rm \NL}$ can be small on WMAP7 scales but large on cluster scales.
  
\begin{figure}
 \epsfxsize=8cm
 \epsffile{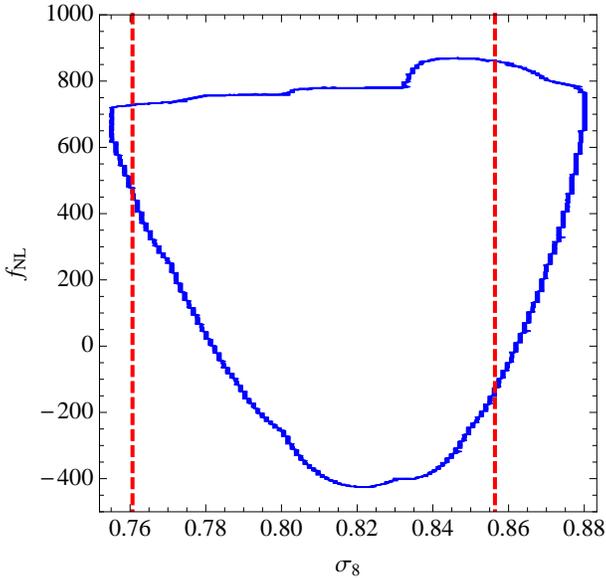}
 \caption{The 95\% posterior probability function contours (solid blue) for WMAP7+BAO+H0 and XMM2235. The 95\% prior  probability function contours from WMAP7+BAO+H0 are plotted as red dashed lines. The $f_{\NL}$ here is on scales much smaller than those constrained by WMAP7 or halo bias measurements.
  \label{posterior}}
\end{figure}

\begin{figure}
 \epsfxsize=8cm
 \epsffile{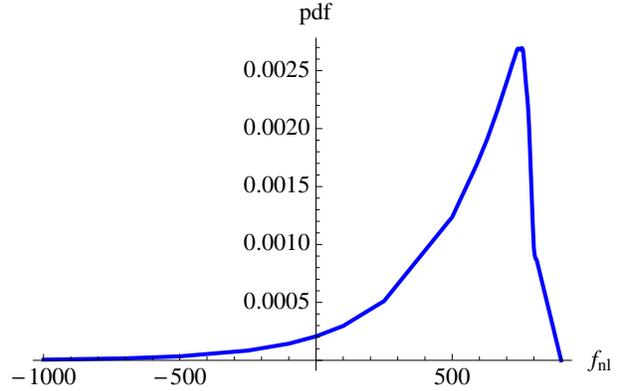}
 \caption{The marginalized posterior probability density function function for $f_{\NL}$  on cluster scales.
 \label{marginalized}}
\end{figure}

 The WMAP7 data  (Larson et al 2010) constrains $\sigma_{8}$ and the other 5 cosmological parameters ($\Omega_{m}, \Omega_{b}, h, n, \tau$) of the flat $\Lambda$CDM model to very high precision. Even tighter constraints are obtained when WMAP7 is combined with baryon oscillation (BAO) and Hubble expansion data (H0)
 (Komatsu et al. 2010). The marginalized distributions of the 6 parameters are very close to Gaussian with small degeneracies between the parameters. 
 Also, as the constraint on $\sigma_{8}$ is much tighter than the one obtainable from 
 XMM2235,  we can approximate the WMAP7+BAO+H0 as a Gaussian prior on $\sigma_{8}$ with mean of 0.809 and a standard deviation of 0.024. The posterior for $f_{\NL}$ and $\sigma_{8}$ is obtained by multiplying the likelihood and prior. The 95\% confidence interval contours are plotted in Fig.~ \ref{posterior}.
As can be seen by comparing Fig.~\ref{likelihood} with Fig.~\ref{posterior}, there is some tension between the higher values of $\sigma_{8}$ preferred by XMM2235 and those favoured by WMAP7 even when a non-zero $f_{\NL}$ is allowed. It can also be seen that WMAP7 completely breaks the degeneracy between $\sigma_{8}$ and $f_{\NL}$. The marginalized posterior probability density function function for $f_{\NL}$ is plotted in Fig.~\ref{marginalized}.
We found that the marginalized probability that $f_{\NL}>0$ was 94\%, the mean of $f_{\NL}$ was
529 and the standard deviation 194.
%The 95\% confidence interval was 
%$-170 \leq f_{\NL}\leq 790$.

We can combine the WMAP7 constraint $f_{\NL}=32\pm21$ (Komatsu et al. 2010) at wave-numbers of about 0.04~Mpc$^{-1}$ to the one obtained here $f_{\NL}=529\pm194$ at wave-numbers of about 0.1~Mpc$^{-1}$ to constrain the non-Gaussian spectral index, defined by
\begin{equation}
f_{\NL}=f_{\NL}^{\rm CMB}\left({k\over k_{\rm CMB}}\right)^{n_{\rm NG}}\, .
\label{nng}
\end{equation}
In order to use Eq.~\ref{nng}, we have to assume that $f_{\NL}$ is the same sign on all scales.
Using the CMB and cluster mean constraints on $f_{\NL}$ implies ${n_{\rm NG}}\sim 3$ which is quite high as it is generally first order in the slow roll parameters, but larger values are possible (Byrnes et al. 2009). Alternatively, 
such a rapid change in $f_{\rm \NL}$ could be explained by
 models that produce a step-function like change in $f_{\rm \NL}$   (Riotto and Sloth, 2010).

%As the non-Gaussianity enhancement of cluster abundance only depends on the skewness, the constraints on the local form of non-Gaussianity, given by Eq.~\ref{fnl}, can be converted to those on 
%the equilateral form of non-Gaussianity by a simple multiplicative factor (Jimenez \& Verde 2009). Using this, our equilateral non-Gaussianity  constraint from XMM2235 is $f_{\NL}=1528 \pm 971$.

\section{Conclusions} 
\label{conclusions}
We found constraints of $f_{\NL}=529\pm 194$ at wave-numbers of about 0.1 Mpc$^{-1}$ using the observations of cluster XMM2235. We found $p(f_{\NL}>0|$XMM2235$)\approx 0.94$ and so XMM2235 is consistent with $f_{\NL}=0$. The WMAP7 data was used to break the degeneracy with $\sigma_{8}$.  In order to do the analysis, we assumed  XMM2235 was the most massive cluster of the survey. If this turns out not to be the case, our results will be conservative as a larger cluster mass would imply more preference for a higher value of $f_{\NL}$. 

As done in Jee et al. (2009), we based our analysis on the $\Lambda$CDM Jenkins et al. (2001) mass function
fitted for the 324 overdensity halo definition. We also tried  several other mass functions.
The mass of XMM2235 can be evaluated for other overdensities by assuming a NFW profile (Hu \& Kravtsov 2003). We found similar results for the 180  overdensity mass functions of Jenkins et al. (2001) and Sheth \& Tormen (1999). It is a concern that in general the mass functions need to be extrapolated to some extent for 
the very high masses we  are considering. This is partly due to the rareness of such massive halos
in the $f_{\NL}=0$ case. It would be useful if this range could be mapped out better by a large number of low resolution N-body simulations.

Our results are based on the assumption, that Matarrese et al. (2000) approach for accounting for the enhancement of cluster  numbers is valid up to $f_{\NL}\sim 1000$, which should be tested against N-body simulations. Recent work (Enqvist et al.\ 2010)
has shown that the high $f_{\NL}$ cutoff in our marginalized distribution posterior distribution of $f_{\NL}$ may be too drastic. This would imply that our results are conservative and that our mean $f_{\NL}$ is higher.
  
Also, selection effects could reduce the 
significance of our result in that rather than using the 11 deg$^2$   one should take the sky area of all surveys 
which have sufficient depth to see clusters at $z\ge  1.4$.
For example if we take the 
the Hoyle et al (2010)  compilation of high-$z$ X-ray cluster surveys which covers an area of about 300 square degrees that would reduce our value of  $p(f_{\NL}>0|$XMM2235$)$ to about 
15\% although some of those surveys do not have good completeness for such a high redshift. Taking into account Sunyaev-Zeldovich surveys may reduce the significance further depending on whether they have sufficient depth to have seen clusters at $z\ge 1.4$. Additionally one could use a larger range of observed high-z clusters to constrain $f_{\NL}$. But a fuller analysis may require a detailed study using the selection functions of all current cluster surveys, so we leave this for future work.

There are several other mechanisms, that could lead to unusually high mass clusters, such as  features in the primordial power spectrum (Chantavat et al. 2008), or
modifications in the growth function due to dynamical dark energy or non-Einstein gravity.
Our results are consistent with $f_{\NL}=0$  but the allowed range of $f_{\NL}$ is quite high. If our central values turn out to be correct, then future observations should be able to detect the non-Gaussianity and its running (LoVerde et al. 2008, Jimenez and Verde 2009, Sefusatti et al. 2009).

\section*{Acknowledgements}
We thank James Jee  for helpful discussions.
LC and CG acknowledge financial support from the Royal Society International Travel Grant 2009/R4. CG is supported by the Beecroft Institute for Particle Astrophysics and Cosmology.

\end{document}